# Nanomagnetic Engineering of the Properties of Domain Wall Atom Traps

T.J. Hayward[1], A.D. West[2], K.J. Weatherill[2], T. Schrefl[3], I.G. Hughes[2] and D.A. Allwood[1]

*[1]Department of Materials Science and Engineering, University of Sheffield, Sheffield, UK*

*[2]Department of Physics, University of Durham, Durham, UK*

*[3] St Pölten University of Applied Sciences, St Pölten, Austria*

We have used the results of micromagnetic simulations to investigate the effects of nanowire geometry and domain wall magnetization structure on the characteristic parameters of magnetic atom traps formed by domain walls in planar ferromagnetic nanowires. It is found that when traps are formed in the near-field of a domain wall both nanowire geometry and wall structure have a substantial effect on trap frequency and adiabaticity. We also show that in certain regimes a trap's depth depends only on the amplitude of an externally applied rotating magnetic field, thus allowing it to be tuned independently of the trap's other critical parameters.

## Introduction

Contemporary laser cooling techniques allow clouds of atoms to be routinely prepared with temperatures in the micro-Kelvin range. In such conditions atom clouds represent idealised quantum mechanical systems that not only allow insight into fundamental phenomena, such as the behaviour of quantum degenerate matter [1,2], but also have great technological potential through the development of matter wave interferometry [3], novel sensors [4] and in information processing [5,6].

To realise these applications atom clouds must not only be confined in velocity space, but also trapped physically. This can be achieved using either optical interactions [7-10] or, for paramagnetic atoms, magnetic interactions [11]. In the latter case atoms are subject to magnetic field gradients created from either current carrying conductors [e.g. 12-14], or ferromagnetic patterns/microstructures [e.g. 15-20]. The possibility of miniaturising these systems and integrating them into substrate-bound "atom chips" [21] makes such approaches extremely attractive for technological applications.

In a previous publication we have used micromagnetic simulations to demonstrate the feasibility of creating atom traps using the monopole-like magnetic fields emanating from domain walls (DWs) in planar magnetic nanowires [22]. These DWs have particle-like properties and can be transported controllably around complex nanowire networks [e.g. 23]. The resulting mobility of atom traps created by such DWs is non-typical for magnetic atom traps based on patterned magnetic microstructures and is extremely attractive for quantum information processing applications. Furthermore, DW atom traps will allow tight confinement of atoms in all three dimensions. Recently we have demonstrated substantial progress towards our goal of experimentally realising DW atom traps by using an array of DWs in nanowires to create a reconfigurable "atom mirror" [24]

While we have previously established the basic feasibility of trapping ultra-cold atoms using a DW, the dependence of trap parameters on the specific properties of the nanomagnetic system forming it has not yet been explored. Here we address this issue by using the results of micromagnetic simulations to investigate how the critical parameters of a DW atom trap, namely its depth, frequency and adiabaticity, depend on both nanowire geometry and the internal magnetisation structure of the DW. The direct effects of these parameters are isolated from those due to variation in the DWs' net monopole moment by maintaining a constant nanowire cross-sectional area throughout our calculations.

Our results show that for traps formed above a DW, nanowire geometry substantially alters an atom trap's frequency, adiabaticity and the maximum obtainable trap depth. We also show that in certain, physically realisable, regimes the trap depth is dependent only on the magnitude of the externally applied rotating magnetic field that is used to ensure that the trap has a non-zero field minimum. This effectively allows the trap frequency and adiabaticity to be "tuned" independently from the trap depth. In combination, these properties are likely to be useful for designing DW atom traps that can be experimentally realised.

## Theoretical Description

The magnetic shape anisotropy of a planar nanowire confines its magnetisation to lie along its length, and hence DWs represent regions of either converging magnetisation (Head-to-Head, H2H) or diverging magnetisation (Tail-to-Tail, T2T) (see figure 1(a)). Each DW therefore carries a net monopole moment with effective "magnetic charge" $q = \pm 2\mu_0 M_s wt$ [24,25], where $M_s$ is the nanowire's saturation magnetisation, *w* and *t* are the nanowire's width and thickness, and $\mu_0$ is the permeability of free space. In previous publications [25, 26] we have shown that the field emanating from a DW can be approximated by assuming that this charge acts as a point-monopole at the DW centre, resulting in the coulomb-like magnetic field:

$$\mathbf{B}_{DW}(\mathbf{r}) = \frac{q}{4\pi r^2}\hat{\mathbf{r}} \qquad \text{(equation 1)}$$

where *r* is the distance of a point from the DW centre. While highly accurate in the far-field, the finite spatial distribution of magnetic poles within the DW causes substantial deviations from this model when considering points closer to the nanowire [26]. This spatial distribution is modified as the geometry of the nanowire and DW magnetisation structure are altered, leading to the dependence of trap parameters on nanowire geometry that we will present later. In this paper the point-monopole model will be used to provide the reader with an intuitive picture of how DW atom traps are formed and as a reference through which to understand how the finite spatial distribution of poles within a DW affects the atom trap it creates. From this point onwards $\mathbf{B}_{DW}$ will be used generically to refer to the field from a DW, rather than solely that calculated using the point-monopole model.

Figure 2(a) illustrates schematically how a magnetic field with form similar to $\mathbf{B}_{DW}$ can be manipulated to create an atom trap. A paramagnetic atom moving adiabatically in a magnetic field gradient $\nabla|\mathbf{B}|$ will experience a force $\mathbf{F} = -m_F g_F \mu_B \nabla|\mathbf{B}|$, where $m_F$ is the atom's magnetic quantum number, $g_F$ is the Landé g-factor and $\mu_B$ is the Bohr magneton. Atoms in states where $m_F g_F > 0$ are attracted to minima in the magnetic field, and are termed "weak–field-seeking". In this paper we will consider $^{87}$Rb atoms that have been optically pumped into the weak-field seeking $5^2S_{1/2}$ $F = 2$, $m_F = 2$ ($g_F = 1/2$) state.

Because $|\mathbf{B}_{DW}|$ increases as distance to the DW decreases, in isolation a DW will simply repel weak-field seeking atoms. To create a field minimum that may be used to trap atoms, an external magnetic field, $\mathbf{B}_{DC}$, is applied in opposition to the dominantly z-axis orientated field directly above the DW. At some height, $z_{trap}$, $\mathbf{B}_{DC}$ exactly cancels $\mathbf{B}_{DW}$ yielding the required field minimum (Fig. 2(b)).

To achieve tight traps with long lifetimes it is required that the magnetic field at the trap center, $|\mathbf{B}|_{min} > 0$. If this criterion is not met atomic states with different $m_F$ become degenerate at the trap minimum, allowing the atoms to perform Majorana spin flips to untrapped states and being lost from the trap. To overcome this problem we consider the Time Orbiting Potential (TOP) approach [27] and apply a rotating magnetic field in the x-y plane, $\mathbf{B}_{TOP}(t)$. Providing that the frequency, $\omega$, of $\mathbf{B}_{TOP}(t)$ is high enough the atoms will only experience a time-averaged field landscape with minimum $|\mathbf{B}_{min}| = |\mathbf{B}_{TOP}|$.

With the application of $\mathbf{B}_{DC}$ and $\mathbf{B}_{TOP}$ the instantaneous components of the magnetic field at a time, *t,* are:

$$\mathbf{B}(\mathbf{r},t) = \begin{pmatrix} B_x(\mathbf{r},t) \\ B_y(\mathbf{r},t) \\ B_z(\mathbf{r}) \end{pmatrix} = \mathbf{B}_{DW} + \mathbf{B}_{DC} + \mathbf{B}_{TOP} = \mathbf{B}_{DW} + \begin{pmatrix} 0 \\ 0 \\ |\mathbf{B}_{DC}| \end{pmatrix} + \begin{pmatrix} |\mathbf{B}_{TOP}|\cos(\omega t) \\ |\mathbf{B}_{TOP}|\sin(\omega t) \\ 0 \end{pmatrix} \qquad \text{(equation 3)}$$

where $\mathbf{B}_{DW}$ is calculated either from the point monopole model or from more complex analytical [26] or numerical [22] models of domain wall pole distributions, and $\omega$ is the angular frequency of $\mathbf{B}_{TOP}$. Assuming that the atoms follow the field adiabatically and experience only the time averaged-field they will then be subject to a magnetic field-landscape defined by:

$$|\mathbf{B}(\mathbf{r})| = \frac{1}{T}\int_0^T |\mathbf{B}(\mathbf{r},t)|\, dt \quad \text{(equation 4)}$$

where $T = \frac{2\pi}{\omega}$. This integral does not have an analytic solution even for the simple monopole model, and hence all of the calculations of $|\mathbf{B}(\mathbf{r})|$ presented in this paper were performed numerically. We note that our approach here assumes that the DW remains fixed as $\mathbf{B}_{\mathrm{TOP}}$ rotates. In reality the larger TOP fields considered in this paper might be sufficient to induce DW motion, however this could be easily prevented by patterning artificial defects into the nanowires.

Having established the basic form of the fields that are used to create a DW atom trap we now turn attention to the critical parameters that define an atom's interaction with a trap.

The trap depth, $U$, defines the minimum energy barrier an atom must overcome to escape from an atom trap, and therefore has a strong influence on an atom's average lifetime within the trap. In general, the depth of a magnetic atom trap can be calculated using $U = m_{\mathrm{F}} g_{\mathrm{F}} \mu_{\mathrm{B}} \left( |\mathbf{B}|_{\infty} - |\mathbf{B}|_{\mathrm{min}} \right)$, where $|\mathbf{B}|_{\infty}$ is the field far from the DW, outside of the trapping potential. In this paper we express $U$ as an effective trap temperature $T = \frac{\Delta E}{k_{\mathrm{B}}}$, where $k_{\mathrm{B}}$ is Boltzmann's constant.

For DW atom traps $|\mathbf{B}|_{\infty}$ is purely due to the isotropic externally applied fields and hence $|\mathbf{B}|_{\infty} = \sqrt{|\mathbf{B}_{\mathrm{DC}}|^2 + |\mathbf{B}_{\mathrm{TOP}}|^2}$. The trap depth is therefore given by:

$$T_{\infty} = \frac{1}{k_{\mathrm{B}}} m_{\mathrm{F}} g_{\mathrm{F}} \mu_{\mathrm{B}} \left( \sqrt{|\mathbf{B}_{\mathrm{DC}}|^2 + |\mathbf{B}_{\mathrm{TOP}}|^2} - |\mathbf{B}_{\mathrm{TOP}}| \right) \qquad \text{(equation 5)}$$

For a TOP trap the calculation of trap depth is complicated by the addition of a second route via which the atoms may escape. While $\mathbf{B}_{\mathrm{TOP}}$ removes the field-zero at the trap centre in the time-averaged field landscape, as it rotates it creates a circle of instantaneous field zeros in the x-y plane below the trap. If an atom encounters this "circle of death" there is a high chance of it performing a spin-flip to an untrapped state and therefore being lost. By considering that for any point on the circle $\sqrt{B_{\mathrm{x}}(\mathbf{r},t)^2 + B_{\mathrm{y}}(\mathbf{r},t)^2} = |\mathbf{B}_{\mathrm{TOP}}|$ and $|B_{\mathrm{z}}(\mathbf{r})| = |\mathbf{B}_{\mathrm{DC}}|$ it can be shown that the time-averaged strength of the magnetic field at any position on the "circle of death" is given by:

$$|\mathbf{B}|_{circle} = \frac{4|\mathbf{B}_{\mathrm{TOP}}|}{\pi} \qquad \text{(equation 6)}$$

The trap depth, $T_{\text{circle}}$, due to this effect is therefore given by:

$$T_{\text{circle}} = m_{\text{F}} g_{\text{F}} \mu_{\text{B}} \frac{(4-\pi)}{\pi k_{\text{B}}} |\mathbf{B}_{\text{TOP}}| \qquad \text{(equation 7)}$$

In the above we assume that $|\mathbf{B}(\mathbf{r})|$ always increases consistently along all vectors between the trap centre and the "circle of death". We have found that this is the case for all of the traps considered in this paper, however we observe empirically that for $|\mathbf{B}_{\text{TOP}}| > 2|\mathbf{B}_{\text{DC}}|$ a field maximum may be found prior to reaching the circle. This is likely to affect the trap depth in cases where large values of $\mathbf{B}_{\mathbf{TOP}}$ are used, or when a trap is formed far from the DW.

In practice the trap depth is limited by whichever of $T_{\text{circle}}$ and $T_{\infty}$ is lower. As $T_{\text{circle}}$ increases with increasing $|\mathbf{B}_{\text{TOP}}|$, while $T_{\infty}$ decreases, the trap depth will be limited by $T_{\text{circle}}$ at low $|\mathbf{B}_{\text{TOP}}|$ and by $T_{\infty}$ at higher $|\mathbf{B}_{\text{TOP}}|$, with a crossover between the two depths occurring at:

$$|\mathbf{B}_{\text{TOP}}|^{\text{cross}} = \frac{\pi}{\sqrt{(16-\pi^2)}} |\mathbf{B}_{\text{DC}}| \quad \text{(equation 8)}$$

As $T_{\text{circle}}$ is only dependent on the value of $|\mathbf{B}_{\text{TOP}}|$ this leads to the remarkable conclusion that for low values of the TOP field the trap depth is entirely independent of the monopole moment of DW that is used to form the trap. It should be noted that this analysis of $T_{\text{circle}}$ was not included in our previous study of DW atom traps [22].

A second important characteristic parameter of an atom trap is the trap frequency, $\omega_{\text{trap}}$, which determines the spatial confinement of trapped atoms, and also the spacing of energy levels within the trap. Treating the trap as a quantum harmonic oscillator (i.e. a quadratic potential):

$$\omega_i = \sqrt{\frac{\mu_{\text{B}}}{m_{\text{A}}} \frac{d^2|\mathbf{B}|}{dr_i^{\,2}}} \qquad \text{(equation 9)}$$

$$\omega_{\text{trap}} = \sqrt[3]{\omega_{\text{x}} \omega_{\text{y}} \omega_{\text{z}}} \qquad \text{(equation 10)}$$

where $\omega_i$ is the characteristic frequency of the trap along Cartesian axis $r_i$ and $m_{\text{A}}$ is the mass of the traped atom. In general the potential landscapes created by DW atom traps are not purely quadratic, and hence the value of $\omega_{\text{trap}}$ best representing the system depends to some degree on the temperature of the trapped atoms. In this paper we simplify this problem by fitting values of the trap frequency over a fixed distance of ± 100 nm from the trap centre. $\omega_{\text{trap}}$ is a particularly critical parameter for TOP traps, as for trapped atoms to respond to the time-averaged magnetic field the TOP field must be rotated at a frequency much greater than the characteristic trap frequency (i.e. $\omega_{\mathrm{trap}} << \omega$). This criteria represents a significant consideration in the design of DW atom traps, as creating large time-orbiting fields at high frequencies is technically challenging, particularly in an ultra-high-vacuum environment.

An important parameter related to $\omega_{\mathrm{trap}}$ is the trap's adiabaticity, which is defined here as $\omega_{\mathrm{L}} / \omega_{\mathrm{trap}}$, where $\omega_{\mathrm{L}}$ is the minimum value of a trapped atom's Larmor frequency as found at the trap centre:

$$\omega_{\mathrm{L}} = \frac{\mu_{\mathrm{B}} \left| \mathbf{B}_{\mathrm{TOP}} \right|}{\hbar} \qquad \text{(equation 11)}$$

where $\hbar$ is the reduced Plank's constant. As the $\omega_{\mathrm{trap}}$ effectively describes the rate at which the magnetic field changes in the atom's frame of reference, and $\omega_{\mathrm{L}}$ describes the rate at which the atom is able to respond to changes in the field direction, this parameter ultimately dictates whether or not the atom's magnetic moment can follow the trap's varying magnetic field adiabatically. Thus, for successful trapping:

$$\frac{\omega_{\mathrm{L}}}{\omega_{\mathrm{trap}}} >> 1. \qquad \text{(equation 12)}$$

## Method

To investigate how the critical parameters of DW atom traps depend on nanowire geometry we consider three 8.4 μm long $Ni_{80}Fe_{20}$ nanowires with differing widths (*w*) and thicknesses (*t*): Nanowire A: (*w* = 200 nm, *t* = 40 nm), Nanowire B: (*w* = 400 nm, *t* = 20 nm), Nanowire C: (*w* = 800 nm, *t* = 10 nm). As all three nanowires have the same cross-sectional area the total monopole moment of their DWs will be identical, and hence any difference between the traps they create are solely due to the spatial distribution of magnetic poles within the DWs.

Physically appropriate DW structures were generated by relaxing simple bi-domain magnetization configurations in accordance with the Landau-Lifshitz-Gilbert equation using a proprietary finite-element micromagnetic code [28]. Standard parameters were used to represent the material parameters of $Ni_{80}Fe_{20}$ (saturation magnetization, $M_S$ = 860 kA/m, exchange stiffness, $A$ = 13 pJ/m, magnetocrystalline anisotropy constant, $K_1 = 0$). A characteristic mesh size of 5 nm was used in the regions of the nanowires containing the DWs, while larger 20 nm meshes were used in uniformly magnetized regions. In all three nanowire geometries vortex DW structure [29] was energetically favored, although in Nanowire C this was bistable with transverse DW structure, allowing the effect of DW magnetization structure on trap parameters to be investigated. Micromagnetically simulated DW structures for all three nanowires are shown in Figure 1(b).

For each nanowire $\mathbf{B}_{DW}$ was calculated from the quasistatic Maxwell equations using a finite element/boundary element method [30]. Calculations were performed across regular meshes (cell size ≤ 20 nm) that extended at least ±200 nm from the trap centre in each direction and were also large enough to fully contain the "circle of death" for a given value of $|\mathbf{B}_{TOP}|$. The magnetic fields created by the nanowires' end domains were subtracted from $\mathbf{B}_{DW}$ by considering point-monopole charges of magnitude –q/2 placed at the nanowires' ends.

In our calculations we consider traps formed at $z_{trap}$ = 500 nm, 1000 nm and 1500 nm, with values of $|\mathbf{B}_{TOP}|$ between 2 and 10 G. $|\mathbf{B}(\mathbf{r})|$ was calculated by first adding an appropriate value of $\mathbf{B}_{DC}$ to $\mathbf{B}_{DW}$ so as to create a field minimum at the desired height, and then numerically integrating *equation 4* across 50 time-steps to simulate the TOP field-landscape. In each time-step the position of the instantaneous field-zero created by $\mathbf{B}_{TOP}$ was located, so as to allow $\mathbf{B}_{circle}$ to be found. $T_{circle}$ was then calculated using Equation 7. $T_{\infty}$ was calculated using Equation 5. Trap frequencies were estimated by performing quadratic fits to $|\mathbf{B}(\mathbf{r})|$ along lines extending ±100 nm from the trap centre and then using Equation 9.

## Results and Discussion

To illustrate the basic variation of trap depth, trap frequency and adiabaticity with $z_{trap}$ and $|\mathbf{B}_{TOP}|$ we initially present calculations for atom traps formed using Nanowire B ($w$ = 400 nm, $t$ = 20 nm) (Figure 3).

Figure 3(a) shows the calculated variation of trap depth with $|\mathbf{B}_{TOP}|$ and $z_{trap}$. It can be seen that for $z_{trap}$ = 500 nm and 1000 nm the trap depth is limited by $T_{circle}$ for all values of $|\mathbf{B}_{TOP}|$, and hence increases linearly in accordance with Equation 7. That $T_{circle}$ is the critical parameter here can be understood by considering Figure 3(d), which plots the value of $|\mathbf{B}_{DC}|$ required to form a trap as a function of $z_{trap}$. For $z_{trap}$ = 500 nm and 1000 nm the required values of $|\mathbf{B}_{DC}|$ are 39 G and 12 G respectively, leading to values of $|\mathbf{B}|_{TOP}^{cross}$ of 49 G and 15 G. As these values are outside of the range $|\mathbf{B}_{TOP}|$ modeled, the trap depth is always limited by $T_{circle}$. In contrast to this, for $z_{trap}$ = 1500 nm, $|\mathbf{B}_{DC}|$ = 5.6 G, and hence $|\mathbf{B}_{TOP}|^{cross}$ = 7.1 G, leading to a transition from $T_{circle}$ to $T_{\infty}$ within the calculated data. Consequently, the trap depth increases linearly to a maximum ~127 μK and then decreases in accordance with Equation 5. In the calculated data the cross-over between $T_{\text{circle}}$ and $T_\infty$ occurs at a slightly higher value of $\left|\mathbf{B}_{\text{TOP}}\right|$ than is predicted by Equation 8 (indicated by dashed red line). This is due to the finite discretisation of the regular mesh, which results in a slight error in the calculated value of $\left|\mathbf{B}\right|_{\text{circle}}$.

In Figure 4 we illustrate how the position of the "circle of death" is modified by changing the value of $\left|\mathbf{B}_{\text{TOP}}\right|$. It is observed that as $\left|\mathbf{B}_{\text{TOP}}\right|$ increases the circle adopts a larger radius and descends towards the DW. This evolution can be understood by considering the monopole-like field pattern generated by the DW, along with the condition that for any point on the circle the in-plane component of $\mathbf{B}_{\text{DW}}$ must be equal in magnitude to $\mathbf{B}_{\text{TOP}}$, while the z-component must be equal and opposite to $\mathbf{B}_{\text{DC}}$: Considering points in an x-y plane containing the trap centre, the magnitude of the in-plane magnetic field $\left(\sqrt{B_x(\mathbf{r},\mathrm{t})^2 + B_y(\mathbf{r},\mathrm{t})^2}\right)$ increases with distance from the trap centre. Hence as $\left|\mathbf{B}_{\text{TOP}}\right|$ increases the circle must adopt a larger radius. However, this increase in the in-plane field is associated with a decrease in $B_{\mathrm{z}}(\mathbf{r})$ and thus the circle must simultaneously descend towards the DW to maintain a z-component equal and opposite to $\mathbf{B}_{\text{DC}}$.

Figure 3(b) plots the variation of $\omega_{\text{trap}}$ with $\left|\mathbf{B}_{\text{TOP}}\right|$. For all trap heights $\omega_{\text{trap}}$ decreases as $\left|\mathbf{B}_{\text{TOP}}\right|$ increases, approximately halving over the range of TOP fields studied. This trend reflects the manner in which the addition of the TOP field "smoothes" the time-averaged field landscape, resulting in more gently varying field gradients (Figure 5(a)). Varying the trap height produces an even stronger variation of trap frequency. For example, with $\left|\mathbf{B}_{\text{TOP}}\right|$ = 6 Oe a trap formed at $z_{\text{trap}}$ = 500 nm has a trap frequency of $1.26\times10^6$ rads$^{-1}$, while under the same conditions a trap at $z_{\text{trap}}$ = 1500nm has a trap frequency of only $8.89\times10^4$ rads$^{-1}$, over an order of magnitude lower. This variation is the result of the rapid decrease in $\frac{d^2\left|\mathbf{B}(\mathbf{r})\right|}{dr_i^{\,2}}$ with height above the DW (Figure 5(b)).

Figure 3(c) illustrates the variation of the trap's adiabaticity as $\left|\mathbf{B}_{\text{TOP}}\right|$ is varied. This variation is found to be stronger than that of $\omega_{\text{trap}}$, because the decrease of $\omega_{\text{trap}}$ with $\left|\mathbf{B}_{\text{TOP}}\right|$ is complemented by a simultaneous linear increase in $\omega_{\text{L}}$. In combination these features allow the adiabaticity to be altered by several orders of magnitude even with the limited range of values of $\left|\mathbf{B}_{\text{TOP}}\right|$ we consider here. The adiabaticity also decreases dramatically as $z_{\text{trap}}$ increases. However, this is solely due changes in $\omega_{\text{trap}}$, as $\omega_{\text{L}}$ depends only on $\left|\mathbf{B}_{\text{TOP}}\right|$ (Equation 11).

We now turn our attention to the effect of nanowire geometry and DW structure on trap parameters.

Figures 6(a), (d) & (g) compare calculated values of the trap depth for Nanowires A, B and C with those from the point monopole model. As in the calculations for Nanowire B discussed earlier, for $z_{\text{trap}}$ = 500 nm and 1000 nm the trap depth is limited by $T_{\text{circle}}$ in the modeled parameter range. Hence, the trap depth depends only on $|\mathbf{B}_{\text{TOP}}|$ and is independent of the nanowire dimensions (the slight differences between the data at $z_{\text{trap}}$ = 500 nm are again the result of finite mesh discretisation). At $z_{\text{trap}}$ = 1500 nm a transition between $T_{\text{circle}}$ and $T_{\infty}$ is again seen. Here, distinct geometry dependence is observed, with $|\mathbf{B}_{\text{TOP}}|^{\text{cross}}$ occurring at lower $|\mathbf{B}_{\text{TOP}}|$ for wider nanowires. Figure 7 illustrates the reason for this: the finite pole distributions in the micromagnetically simulated nanowires lead to reduced magnetic fields at a given height in comparison to the point-monopole model [26]. This difference grows as the nanowires' widths increase due to the pole-distributions become more extended. Thus, lower values of $|\mathbf{B}_{\text{DC}}|$ are required to create the trap and $|\mathbf{B}_{\text{TOP}}|^{\text{cross}}$ is consequently reduced. In regimes where $T_{\infty}$ dominates the trap depth is also lower for wider nanowires, as the reduced value of $|\mathbf{B}_{\text{DC}}|$ leads to lower values of $|\mathbf{B}_{\infty}|$. An importance conclusion here is that for a given trap height and DW charge, nanowire geometry ultimately determines the maximum trap depth that may be obtained.

Figures 6(b), (e) & (h) present calculated values of $\omega_{\text{trap}}$ for all three nanowire geometries. Decreasing the width of the nanowire can be seen to increase the trap frequency for given height and TOP field, with the point monopole model representing an upper limit of what may be obtained for given DW charge. At $z_{\text{trap}}$ = 500 nm a substantial effect is observed, with Nanowire C producing traps with frequencies ~300% lower than those created by Nanowire A, while at greater values of $z_{\text{trap}}$ the differences between the three geometries are less pronounced. The origin of these effects lies in the fact that, in the near-field, more extended pole distributions result in more slowly varying field gradients than more concentrated pole distributions. Moving towards the far-field, the effects of an extended pole distribution become less relevant, resulting in the convergence of the curves towards that of the point-monopole as $z_{\text{trap}}$ increases. The reader may also note an apparent "flattening" of the curves at low $|\mathbf{B}_{\text{TOP}}|$ when $z_{\text{trap}}$ = 500 nm. This is due to the non-harmonic shape of the trapping potential: Under these conditions the field landscape shows notable deviations from quadratic form over the fitted data range (±100 nm).

Figures 6(c), (f) & (i) illustrate how the atom traps' adiabaticity is modified by nanowire geometry. As $\omega_{\text{L}}$ is independent of nanowire geometry, the observed variation is entirely due to the variation of $\omega_{\text{trap}}$ described in the previous paragraph. Thus, the adiabaticity increases with increasing nanowire width, with this dependence becoming less significant at larger trap heights.

Having discussed the effect of nanowire geometry we now discuss how the internal magnetization structure of a DW affects the parameters of an atom trap. As is well known, planar magnetic nanowires support two basic types of DWs [29]: the “transverse” form, in which the magnetization of the DW lies perpendicular to the nanowires length, and the “vortex” form where the DW magnetization rotates around a nanoscopic “core” of out-of-plane magnetization. As indicated earlier in the paper, vortex DW structure is energetically favorable in all three of the modeled nanowire geometries, however, Nanowire C will also support a meta-stable transverse DW configuration (Figure 1(b)). Studying traps in this nanowire therefore allows the effect of DW magnetization structure to be isolated from that due to total DW charge and nanowire geometry. The results of calculations comparing the characteristic parameters of traps formed by the two DW geometries can be found in Figure 8.

The basic effect of DW structure can be understood by considering that moving from a vortex to a transverse DW effectively compresses the DW pole distribution. The result of this is that transverse walls produce traps with higher values of $\left|\mathbf{B}_{\mathrm{TOP}}\right|^{\mathrm{cross}}$, $T_{\infty}$, and $\omega_{trap}$, and lower values of adiabaticity than an equivalent vortex wall. A further noticeable effect is that, with transverse DWs, traps are no longer formed above the centre of the nanowire as they are for vortex DWs, but are displaced ~200 nm towards the edge of the nanowire. This reflects the symmetry of the triangular transverse DW structure which, as we have shown previously [26,31], results in an offset in the position of the effective centre of the DWs charge distribution.

## Conclusions

In this study the results of micromagnetic simulations have been used to investigate the effect of nanowire geometry and DW magnetization structure on the critical parameters of atom traps formed using DWs in planar magnetic nanowires.

Our results indicate that when considering traps in the near field of a DW (i.e. $z_{\mathrm{trap}}$~w), nanowire geometry has a substantial effect on both trap frequency and adiabaticity, and also modifies the maximum obtainable trap depth. For given total DW monopole moment, the adiabaticity increases with nanowire width, while the trap frequency and maximum depth decrease. These effects can be understood by a broadening of the DW charge distributions as nanowire width increases. As the height of a trap is increased towards the far-field the effect of nanowire geometry becomes less pronounced due to the field from the DW’s charge distribution tending towards the limiting case of that from a point-monopole.

We have also observed differences between traps formed by transverse and vortex walls in a nanowire of the same geometry. For given trap height and DW monopole moment a transverse wall produces a trap with a higher maximum depth and frequency, and a lower adiabaticity than a trap formed by a vortex DW in the same nanowire. Again, these effects become less important as the trap height increases.

While nanowire geometry and DW structure will undoubtedly be useful tools in optimizing trap properties perhaps a more significant result of this work is the observation that for certain regimes of external parameters the trap depth depends only on the magnitude of the TOP field, and is therefore independent of nanowire geometry, DW magnetization structure, total DW monopole moment and trap height. The upshot of this is that these parameters may be used together to "tune" a trap's frequency and adiabaticity while maintaining an experimentally appropriate trap depth. Particularly exciting is the ability to use the trap height in this way due to the strong dependence of trap frequency and adiabaticity upon it. For example, simply by varying nanowire geometry and trap height within the limited ranges considered in this paper, both trap frequency and adiabaticity can be tuned by more than an order of magnitude, while maintaining a trap depth in excess of 100 μK.

In combination the effects we describe in this paper are likely to be extremely useful in the design and optimization of DW atom traps that can be experimentally realized.

The authors thank the Engineering and Physical Sciences Research Council for financial support (Grant Nos. EP/F024886/1 and EP/F025459/1).

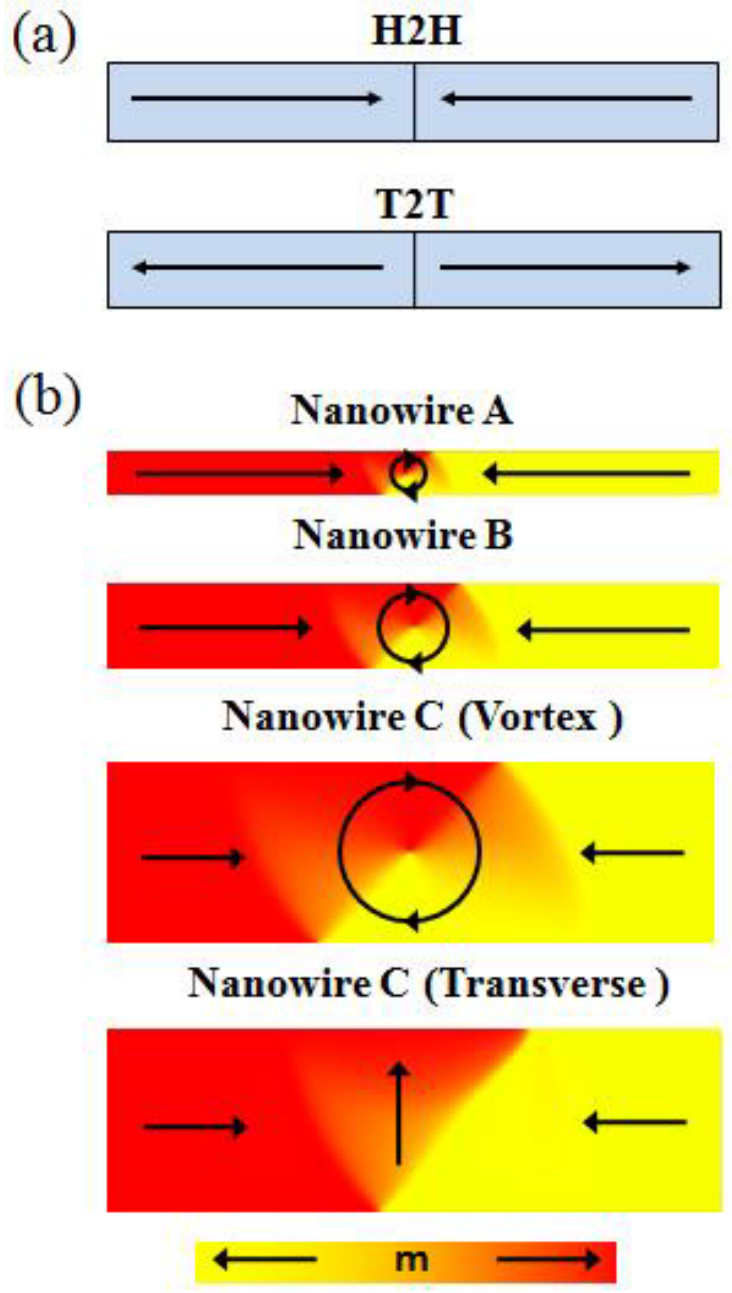

Figure 1: (a) Schematic diagrams of head-to-head (H2H) and tail-to-tail (T2T) domain walls. (b) Micromagnetically simulated domain wall structures for nanowires A, B and C. [COLOR ONLINE]

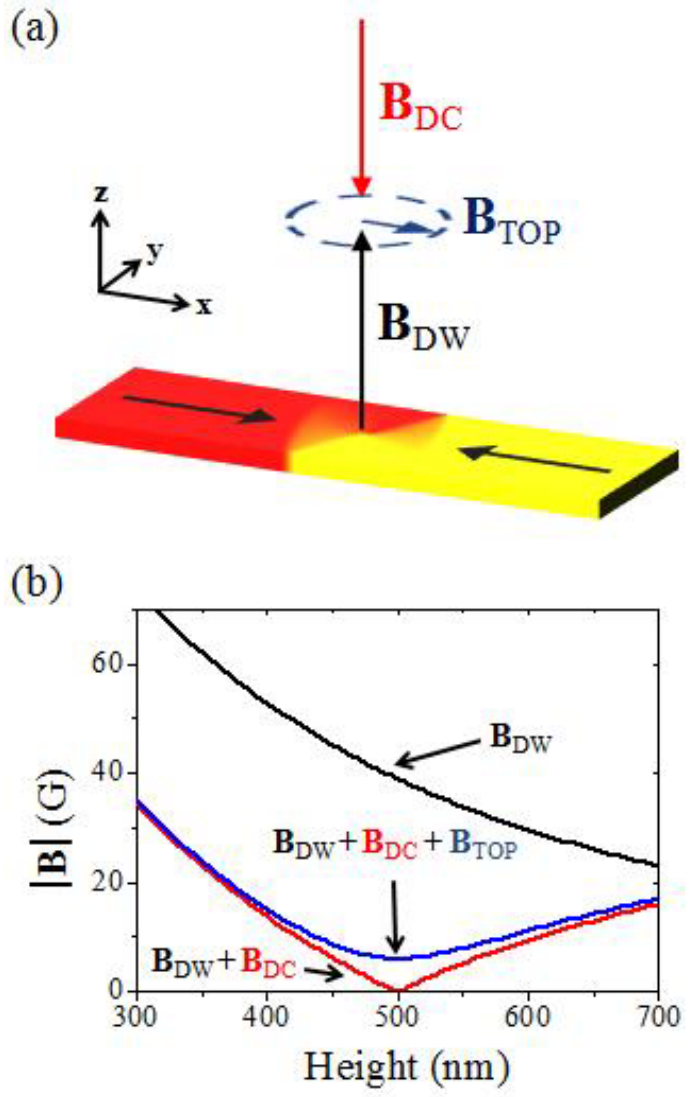

Figure 2: (a) Schematic representation of the experimental geometry required to create a DW atom trap. (b) Plots showing calculated magnetic field as a function of height above a DW in a $Ni_{80}Fe_{20}$ nanowire. The plots illustrate how the externally applied fields ($\mathbf{B}_{DC}$ and $\mathbf{B}_{TOP}$) are combined with $\mathbf{B}_{DW}$ to create an atom trap with a non-zero field minima. Data is shown for Nanowire B (w = 400 nm, t = 20 nm) with the trap formed 500 nm above its centre. $|\mathbf{B}_{TOP}|$ = 6 Oe. [COLOR ONLINE].

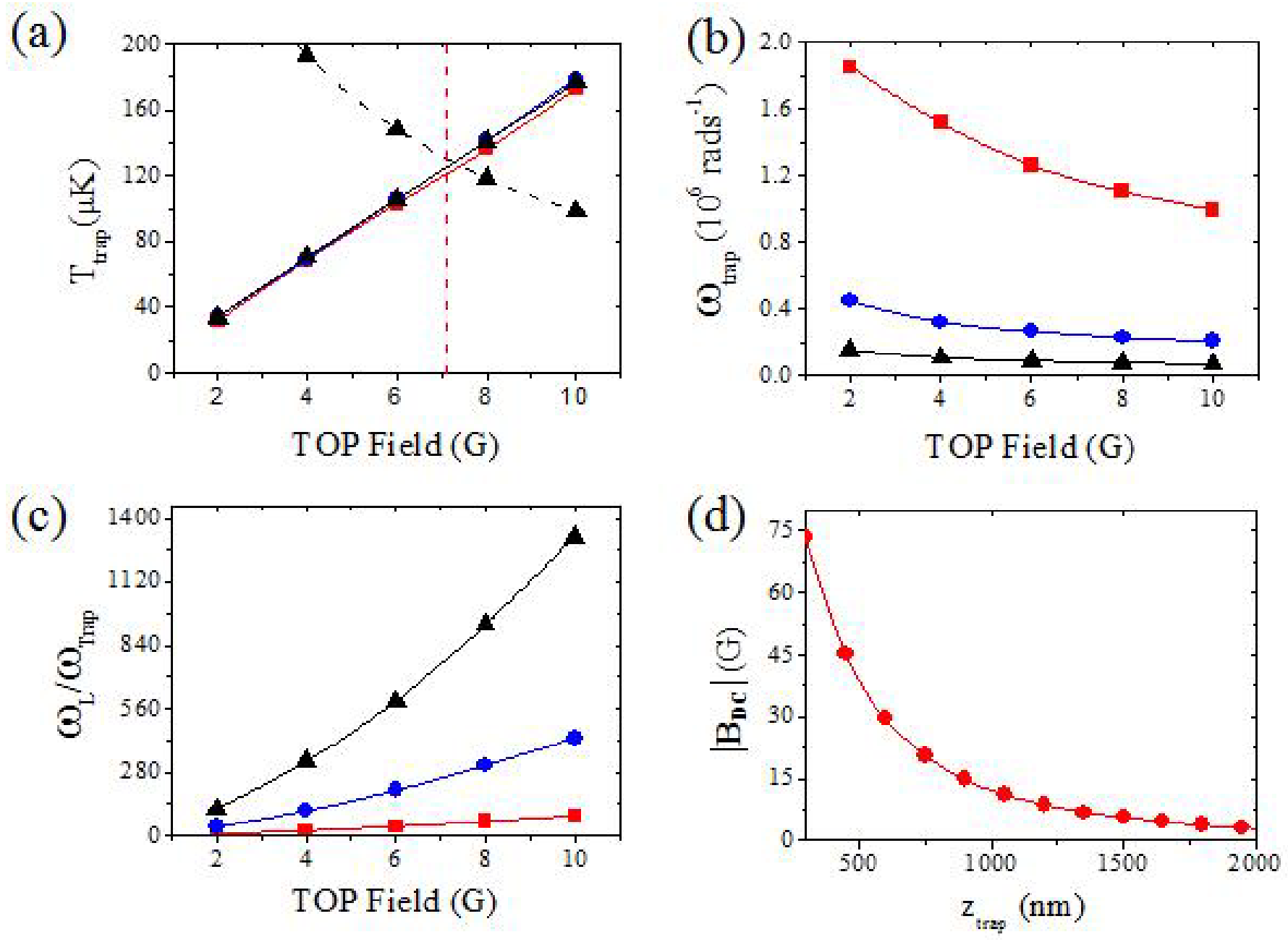

Figure 3: Calculated values of (a) trap depth, (b) trap frequency and (c) adiabaticity as a function of $\left|\mathbf{B}_{\mathrm{TOP}}\right|$ for Nanowire B ($w$=400 nm, $t$ =20nm). Data is shown for $z_{\mathrm{trap}}$ = 500 nm (squares), 1000 nm (circles) and 1500 nm (triangles). In (a) full lines represent $T_{\mathrm{circle}}$ while dashed lines represent $T_{\infty}$. (d) Value of $\left|\mathbf{B}_{\mathrm{DC}}\right|$ required to create a trap as a function of $z_{\mathrm{trap}}$. [COLOR ONLINE].

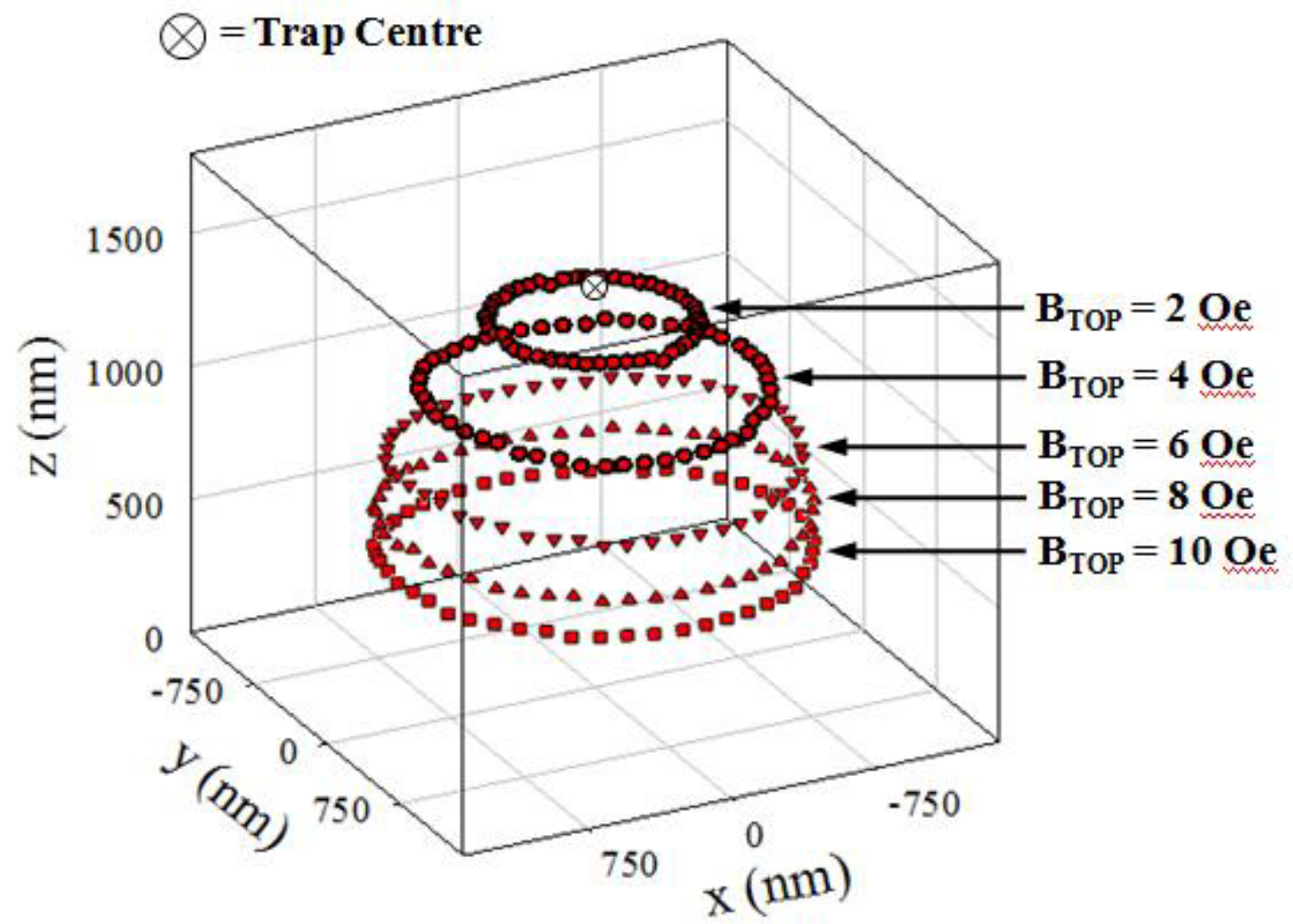

Figure 4: Evolution of the "circle of death" as $\left|\mathbf{B}_{\mathrm{TOP}}\right|$ is varied. Data is shown for Nanowire B with $z_{\mathrm{trap}}$ = 1500 nm. [COLOR ONLINE].

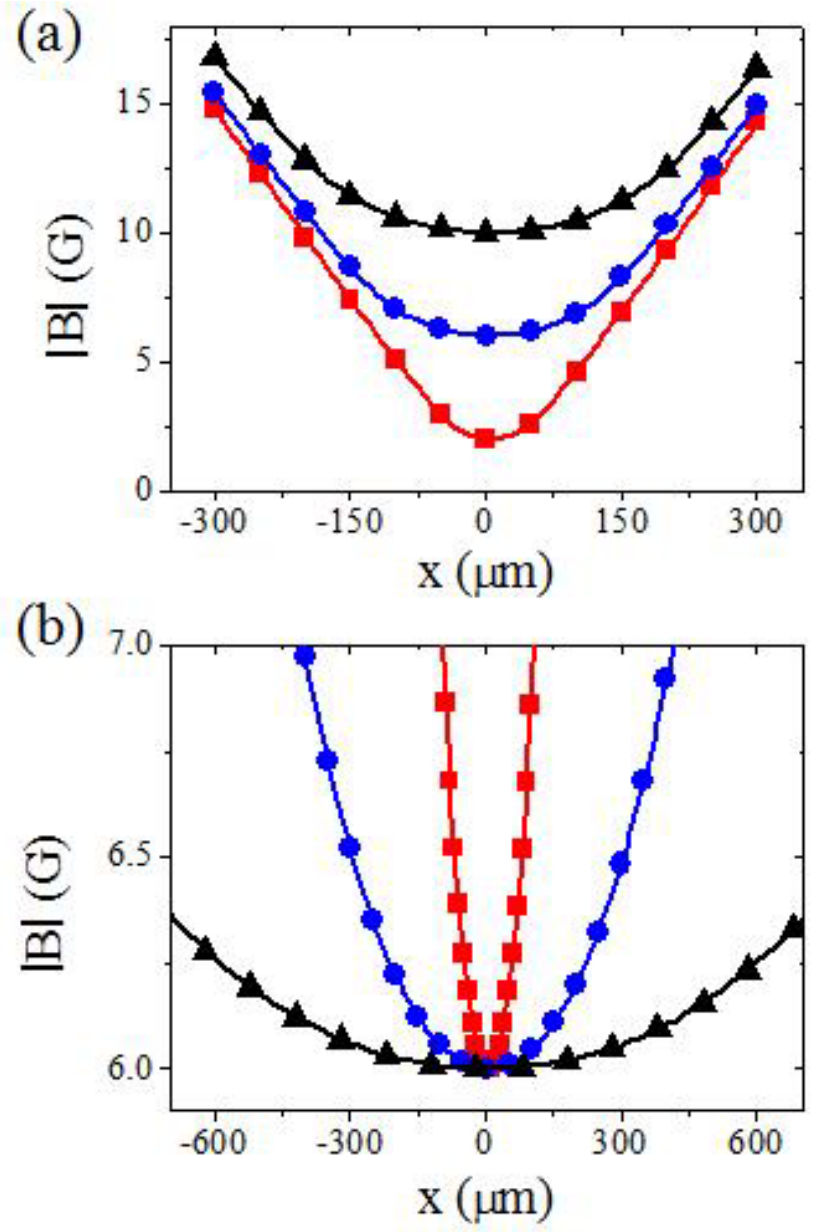

Figure 5: (a) Variation of $\left|\mathbf{B}\right|$ with x position for Nanowire B with $z_{\text{trap}}$ = 1500 nm. Data is shown for $\left|\mathbf{B}_{\text{TOP}}\right|$ = 2 G (squares), 6 G (circles) and 10 G (triangles). (b) Variation of $\left|\mathbf{B}\right|$ with x position for Nanowire B with $\left|\mathbf{B}_{\text{TOP}}\right|$ = 6 G. Data is shown for $z_{\text{trap}}$ = 500 nm (squares), 1000 nm (circles) and 1500 nm (triangles). [COLOR ONLINE].

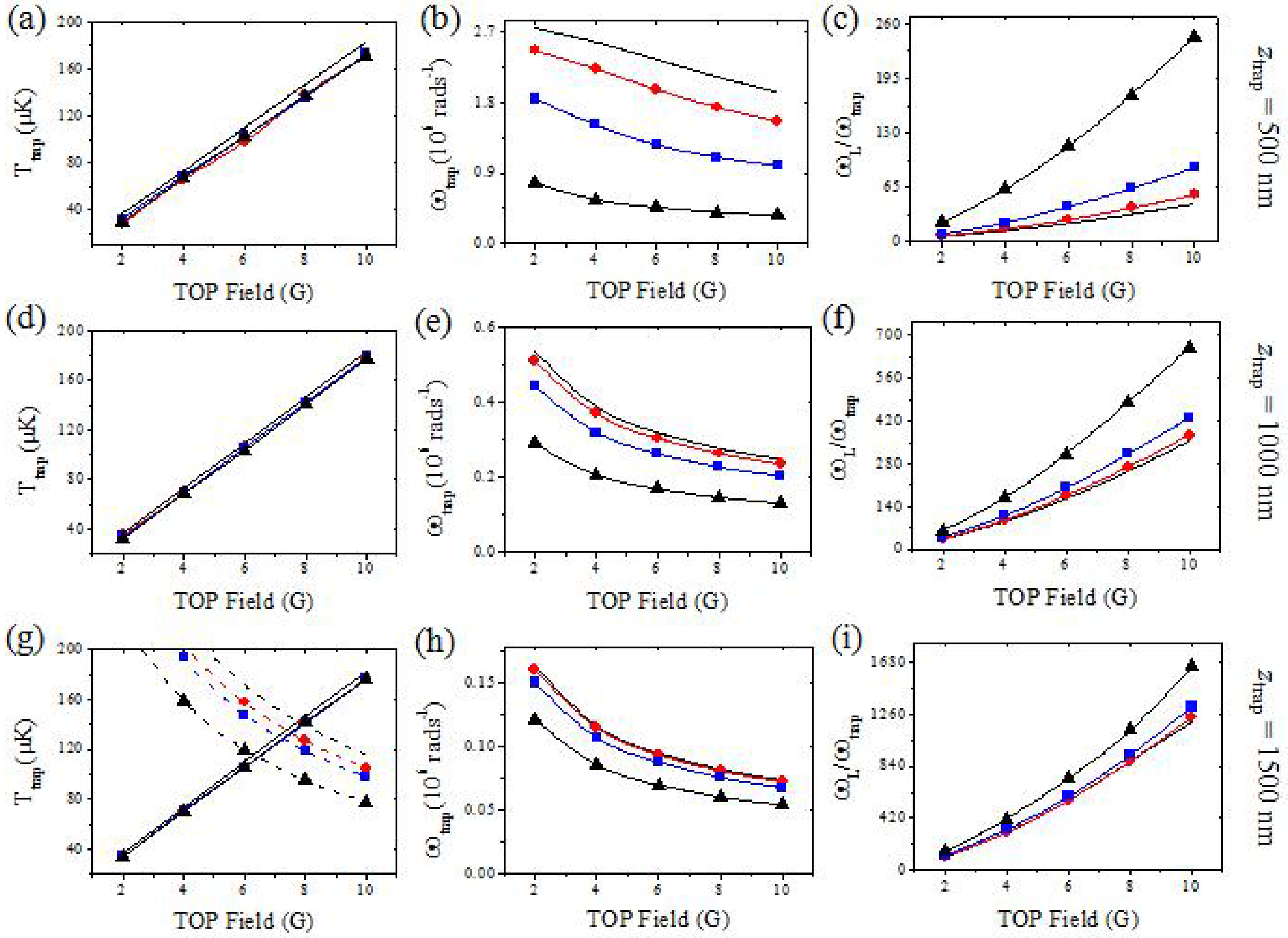

Figure 6: Variation of trap depth, frequency and adiabaticity with nanowire geometry. Data is shown for Nanowires A (circles), B (squares) and C (triangles), as well as for the point-monopole model (no symbols). (a), (b) & (c) $z_{\text{trap}}$ = 500 nm. (d), (e) and (f) $z_{\text{trap}}$ = 1000 nm. (g), (h) and (i) $z_{\text{trap}}$ = 1500 nm. In (a), (d) and (g) full lines represent $T_{\text{circle}}$ while dashed lines represent $T_{\infty}$. [COLOR ONLINE]

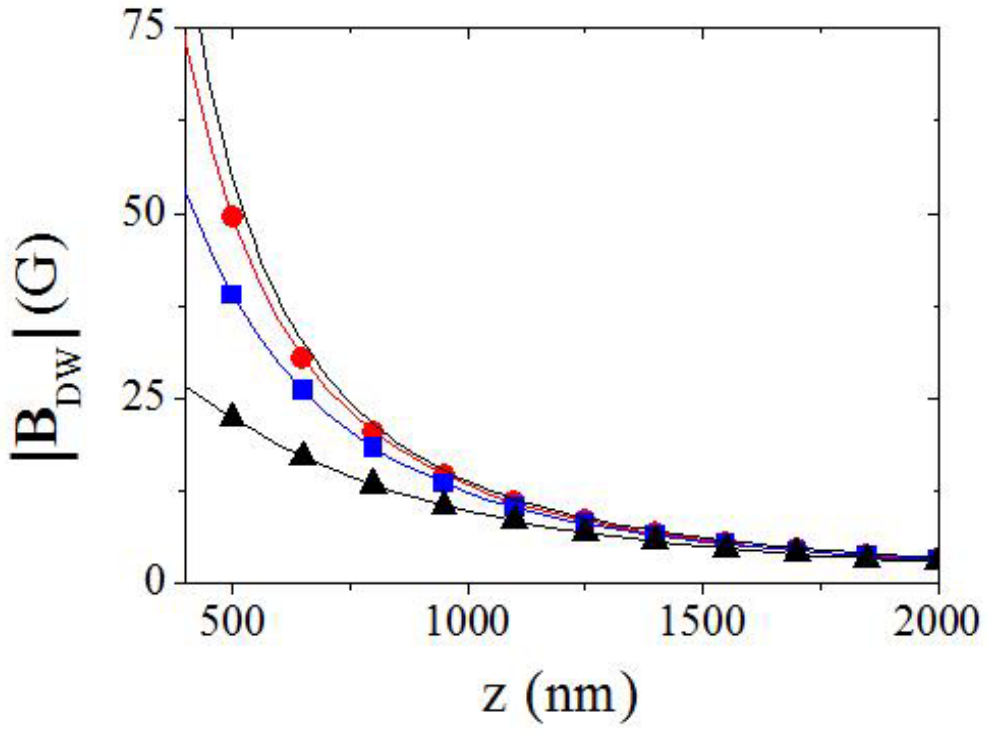

Figure 7: Variation of $|\mathbf{B}_{\text{DW}}|$ with height above centre of DW. Data is shown for Nanowires A (circles), B (squares) and C (triangles), as well as for the point-monopole model (no symbols). [COLOR ONLINE]

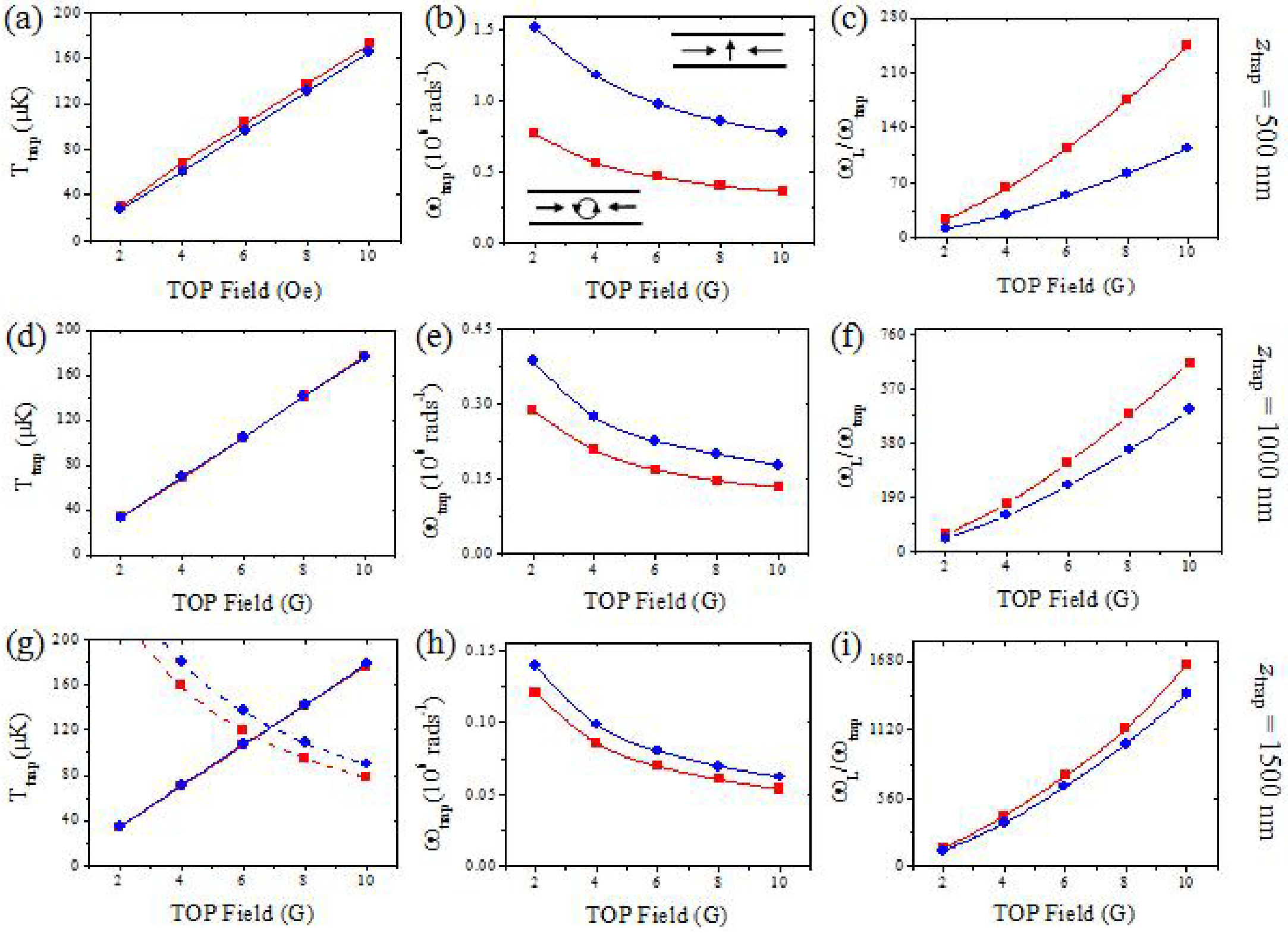

Figure 8: Comparison between the parameters of traps formed by a vortex DW (squares) and a transverse DW (circles) in Nanowire C. (a), (b) & (c) $z_{\text{trap}}$ = 500 nm. (d), (e) and (f) $z_{\text{trap}}$ = 1000 nm. (g), (h) and (i) $z_{\text{trap}}$ = 1500 nm. In (a), (d) and (g) full lines represent $T_{\text{circle}}$ while dashed lines represent $T_{\infty}$. [COLOR ONLINE]